\documentclass[12pt,onecolumn,prd,
tightenlines,eqsecnum,
superscriptaddress,nofootinbib,preprintnumbers, nolongbibliography]{revtex4-2}

\usepackage[utf8x]{inputenc}
\usepackage[T1]{fontenc}
\usepackage[english]{babel}

\usepackage{graphicx}
\usepackage{float}

\usepackage{amsmath}	
\usepackage{amssymb}
\usepackage{amsfonts}
\usepackage{mathrsfs}
\usepackage{mathtools}
\usepackage{amsthm}
\usepackage{upgreek}

\usepackage{braket}
\usepackage{slashed}
\usepackage{hyperref}

\usepackage{xcolor}
\definecolor{unamblue}{cmyk}{1 0.79 0.12 0.59}

\usepackage{color}

\usepackage[]{todonotes}

\usetikzlibrary{
	arrows.meta,
	calc,
	decorations.markings,
	fit,
	matrix,
	positioning,
	shapes
}

\tikzset{
massless/.style={line width=0.2ex},
masslesscollinear/.style={line width=0.3ex, teal, dotted},
bigmassless/.style={line width=0.3ex},
reduced/.style={line width=0.2ex, dashed, dash phase=3pt},
dotdot/.style={line width=0.2ex, dashed, dash phase=2pt},
loop arrow/.style={postaction={decorate, decoration={markings, mark=at position 0.65 with {\arrow{Stealth}}}}}
}

\newcommand{\DBox}{
\includegraphics{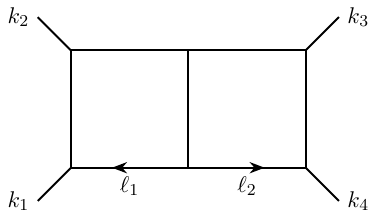}
}

\newcommand{\PentaBox}{
\includegraphics{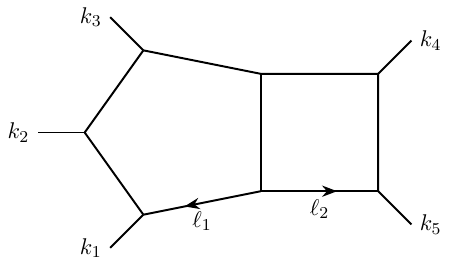}
}

\usepackage[]{hyperref}
\hypersetup{
  colorlinks=true,%
  citecolor=unamblue,%
  filecolor=unamblue,%
	 linkcolor=unamblue,%
	 urlcolor=unamblue,
	 bookmarksnumbered=true,     
	 bookmarksopen=true,         
	 bookmarksopenlevel=1,       
	 pdfstartview=Fit,           
	 pdfpagemode=UseOutlines,
	 pdfpagelayout=TwoPageRight
	}

\def\ellset{{\pmb{\ell}}}
\def\kset{\pmb{k}}

\def\Numer{\mathcal{N}}
\def\Den{\mathcal{D}}
\def\Integ{I}

\newcommand{\dd}{\mathrm{d}}

\def\eps{\epsilon}
\def\Ord{\mathcal{O}}
\def\Lowering#1{\hat{L}^{#1}}

\definecolor{bluebell}{rgb}{0.64, 0.64, 0.82}

\begin{document}
\title{Seedless Reduction of Feynman Integrals} 

\author{Leonardo de la Cruz}
\affiliation{Institut de Physique Th\'eorique, Université Paris-Saclay, CEA, CNRS, F-91191 Gif-sur-	Yvette Cedex, France}
\affiliation{\textsf{\rm\sf leonardo.delacruz@ipht.fr}}

\author{David A. Kosower}
\affiliation{Institut de Physique Th\'eorique, Université Paris-Saclay, CEA, CNRS, F-91191 Gif-sur-	Yvette Cedex, France}
\affiliation{\textsf{\rm\sf david.kosower@ipht.fr}}
\date{\today}
\begin{abstract}
We show how to construct a complete set of lowering
operators, whose successive application reduces an 
arbitrary Fenyman integral to a combination of
master integrals.  The construction builds systems of
equations for generic integral indices using 
IBP-generating vectors.  The solution to each system is
a lowering operator.
\end{abstract}
	\maketitle
\newcommand{\powprop}{a}
\newcommand{\powpropn}{{\bar a}}
\newcommand{\irrepsp}{ \mathscr{I}}
\newcommand{\densp}{ \mathscr{D}}
\newcommand{\repsp}{ \mathscr{R}}
\newcommand{\vectred}{\pmb{a}}
\newcommand{\vectirred}{\pmb{b}}
\def\eqn#1{eq.~\eqref{#1}}
\def\eqns#1#2{eqs.~\eqref{#1} and~\eqref{#2}}
\def\sect#1{Sect.~{\ref{#1}}}

\section{Introduction}

Frontier calculations in quantum field theory typically involve
hundreds, thousands, or even tens of thousands of Feynman integrals.
These integrals are not independent.  There are many linear
relations between them, which can be obtained by writing
down total derivatives.  These relations, known as
integration-by-parts (IBP) identities, reduce the number of
independent integrals to a much smaller number.  
The idea of using IBP identities was introduced by
Chetyrkin and Tkachov~\cite{Tkachov:1981wb,Chetyrkin:1981qh}, 
and systematized by Laporta~\cite{Laporta:2000dsw}.

Solving
the system of IBP relations gives the relations reducing dependent
integrals to basis integrals.  Conceptually, solving the
IBP system is just a matter of performing Gaussian elimination,
but the sizes of systems of interest means that great care
and ingenuity has to be invested in building general-purpose
systems to perform these reductions.  Over the years, 
researchers have written a number of codes to carry out this
task: \textsf{AIR\/}~\cite{Anastasiou:2004vj},
\textsf{FIRE\/}~\cite{Smirnov:2008iw,Smirnov:2025prc}, 
\textsf{Reduze\/}~\cite{Studerus:2009ye,vonManteuffel:2012np}, 
\textsf{LiteRed\/}~\cite{Lee:2012cn, Lee:2013mka}, 
\textsf{Kira\/}~\cite{Klappert:2020nbg, Lange:2025fba}, \textsf{NeatIBP+Kira\/}~\cite{Wu:2025aeg},
and \textsf{Blade\/}~\cite{Guan:2024byi}.    
These implement Laporta's approach with sophisticated improvements.  Researchers have also investigated how to choose a basis of master integrals \cite{Smirnov:2020quc, Usovitsch:2020jrk,e-collaboration:2025frv, Bree:2025tug}. Nonetheless, IBP reductions remain
an important bottleneck to performing amplitude calculations at higher loop order.

Feynman integrals depend on invariants of the external
momenta, and also on the exponents
of the propagators and of numerator factors.  The latter are 
called \emph{indices\/}.
In standard Laporta approaches, one solves systems of 
equations for specific values of indices. 
Moreover, in following Laporta's approach, one of the choices 
to be made
is the initial choice of `seed' integrals, which should make it
efficient to solve for desired target integrals. This
particular choice has been the subject of increased scrutiny
recently~\cite{vonHippel:2025okr, Song:2025pwy, Lange:2025fba,Smirnov:2025prc}.

Some authors have explored recurrence relations for generic exponents;
see for instance ref.~\cite[Chapter 5]{Smirnov:2006ry} and references therein. Smirnov 
made an attempt to use
them in earlier versions of \textsf{FIRE\/} (versions 1--3), 
but abandoned them in later versions as explained in ref.~\cite{Smirnov:2025dfy}. 
Later, Lee made use of these ideas in 
\textsf{LiteRed\/}, obtaining some general rules heuristically and mixing them with a 
Laporta  approach for fixed values of some exponents. 

Codes using the Laporta approach inevitably introduce higher powers of denominators 
\textit{aka\/} `dotted' integrals. In Yang--Mills and gravity theories such integrals are not needed,
and they are typically removed as part of solving the IBP system. 
They can be avoided entirely by making use of special vectors,  known as 
\textit{IBP-generating vectors\/}, in generating the total 
derivatives~\cite{Gluza:2010ws,Schabinger:2011dz, Barakat:2022qlc}. The \textsf{NeatIBP\/}
code~\cite{Wu:2023upw} implements these ideas (using the Baikov 
representation~\cite{Baikov:1996iu}) following earlier ideas~\cite{Larsen:2015ped,Georgoudis:2016wff,Bosma:2017hrk,Bohm:2017qme}.  Ita used these vectors in developing a formalism for separating surface terms~\cite{Ita:2015tya}, and then Abreu \textit{et al.\/}  used them for numerical unitarity at higher loops~\cite{ Abreu:2017idw,Abreu:2017xsl, Abreu:2017hqn,Abreu:2018jgq, Abreu:2018zmy,Abreu:2020xvt}.

More recently, 
Smith and Zeng~\cite{Smith:2025xes} have made use of IBP-generating vectors 
to develop an improved set of 
rules for generic exponents alongside use of a standard
Laporta approach for specific values of some exponents. Liu 
and Mitov~\cite{Liu:2025udl} have considered a different approach to developing rules for 
generic exponents. Feng, Li, Liu, Ma, and Zhang~\cite{Feng:2025leo} and
Chen, Feng, and Zhang~\cite{Chen:2025gqu} have 
developed reduction rules for generic exponents based on 
generating functions and differentiation. 
Parallel 
developments~\cite{Mastrolia:2018uzb,Frellesvig:2019kgj,%
Frellesvig:2019uqt,Weinzierl:2020xyy,Frellesvig:2020qot,%
Brunello:2023rpq,Brunello:2024tqf} use 
intersection theory to simplify Feynman integrals.
Page and Song~\cite{Page:2025gso} studied the connection
between intersection theory and IBP-generating vectors;
Coro, Novichkov, Page, and Song~\cite{Coro:2025kha} 
then built on this to explore integral reduction based on
Landau analysis.

In this article, we investigate an approach different from the Laporta one,
in which seeds are avoided entirely.  
We develop a complete set of
operators that reduce integrals with generic 
indices to simpler integrals.  Repeated application of the rules will reduce an
arbitrary integral to a combination of basis integrals. We rely on IBP-generating vectors as a key part of our construction.
This follows part of the spirit of work by
one of the authors~\cite{Kosower:2018obg}, which sought to find closed-form
solutions for one class of numerators. 

This paper is organized as follows. We present the general setup and approach in Sect.~\ref{SetupSection}. 
In Sect.~\ref{DoubleBoxSection}, we illustrate our approach with the massless double box. 
We consider the pentabox in Sect.~\ref{PentaboxSection}. We discuss some implementation considerations
in Sect.~\ref{ImplementationSection}, and give our conclusions in Sect.~\ref{Conclusions}. 

\def\ISP{J}
\def\nvector{n_v}
\def\nisp{{\bar c}}
\def\ndenom{c}

\section{Setup and Strategy} 
\label{SetupSection}
 We consider dimensionally regulated Feynman integrals in $D$ space-time dimensions,
 \begin{equation}
    \Integ[\Numer(\ellset,\kset)] = \int 
      \prod_{j=1}^L \frac{\dd^D\ell_j}{(2\pi)^D} \;
      \frac{\Numer(\ellset,\kset)}
      {\Den_1^{\powprop_1}\cdots \Den_{\ndenom}^{\powprop_{\ndenom}}}\,,
\label{integral-loop-withnumerator}
\end{equation}
where $\ellset$ denotes the $L$ loop momenta, and $\kset$ 
the set
of $n_e$  external momenta, which are strictly 
four-dimensional. 
The denominators have the form,
\begin{align}
 \Den_e = (M_e)^{jr}\ell_j\cdot \ell_r 
     + 2 (Q_e)^{jr} \ell_j\cdot k_r +Y_e +i\varepsilon\,, 
\end{align}
where the matrices $M_e$ and $Q_e$ have dimensions 
$L\times L$ and $L\times (n_e-1)$ respectively.  We use an
additional set of expressions of the same general form which
are independent of the denominators, called irreducible
scalar products (ISPs) and denoted here by $\ISP_i$.  
Along with the denominators and one chosen scalar product $s_0$
of external momenta, these give us a basis for 
expressing any scalar product of external and loop momenta. 
Accordingly, we can express a general 
numerator $\Numer(\ellset,\kset)$
in terms of $\Den$s, $\ISP$s, and $s_0$.  We rewrite other
scalar products $s_{ij}=(k_i+k_j)^2$ as
$\chi_{ij} s_0$ in terms of dimensionless parameters $\chi_{ij}$.
Numerator terms with factors
of any $\Den$ simply give us similar integrals with fewer
denominators, so without loss of generality we can consider
numerators built out of a product of $\ISP$s,
 \begin{equation}
    \Integ[\pmb{\powprop};-\pmb{\powpropn}] = \int 
      \prod_{j=1}^L \frac{\dd^D\ell_j}{(2\pi)^D} \;
      \frac{\ISP_1^{\powpropn_1}\cdots
      \ISP_{\nisp}^{\powpropn_{\nisp}}}
      {\Den_1^{\powprop_1}\cdots 
      \Den_{\ndenom}^{\powprop_{\ndenom}}}\,.
\label{GeneralIntegral}
\end{equation}
We follow the usual convention that indices $\powpropn$
associated to numerator factors take negative values, along
with the usual labeling convention for the integral.  We expect
the approach
described here to be applicable with small modifications
to linearized or eikonal propagators, but we will not consider them in the
present article.

We will make use of IBP-generating vectors~\cite{Gluza:2010ws}.
Each `vector' is an $L$-tuple of Lorentz vectors, designed 
to avoid introducing
doubled propagators into IBP relations.  Denoting the 
loop index
by $A$, each generator vector satisfies,
\begin{equation}
    v_A^\mu\,\frac{\partial\Den_i}{\partial\ell_A^\mu} =
    g_i\Den_i\,, \quad \textrm{no\ sum\ on\ } i\,,
\label{VectorDefiningEquation}
\end{equation}
where the $A$ index is implicitly summed over.  In this constraint,
each $g_i$ is a polynomial in the denominators, ISPs, 
and $s_0$, 
or equivalently in the set of `variables' consisting of
all dot products of loop momenta with themselves, or with a 
basis of $n_b=\min(n_e-1,4)$ external momenta,
\begin{equation}
    \{ \ell_i^2\,, \ell_i\cdot\ell_j\,, \ell_i\cdot k_j\,, s_0\}\,.
\label{Variables}
\end{equation}
Each element of $v_A^\mu$ is a sum over coefficients times basis
vectors,
\def\cdeg{\mathop{\rm cdeg}\nolimits}
\def\deg{\mathop{\rm deg}\nolimits}
\begin{equation}
    v_A^\mu = \sum_{j=1}^L f_{Aj}^{\vphantom{\mu}}\ell_j^\mu + 
              \sum_{j=1}^{n_b} h_{Aj}^{\vphantom{\mu}} b_j^\mu\,.
\end{equation}
The coefficients $f_{Aj}$ and $h_{Aj}$ are polynomials in
the variables~\eqref{Variables}.  We define 
the operator $\cdeg$ to 
extract the degrees of these coefficients in the variables~\eqref{Variables}, 
\begin{equation}
    \cdeg v_A^\mu = \deg(f_{Aj}) = \deg(h_{Aj})\,.
\end{equation}
In this equation, $\deg$ is the degree of its argument in
terms of the variables~\eqref{Variables}.  The polynomials are necessarily 
homogeneous.  (Four basis vectors suffice because as usual
we take all external vectors to be strictly four-dimensional.)

Thanks to \eqn{VectorDefiningEquation}, a typical total derivative
takes the form,
\begin{equation}
    \frac{\partial}{\partial\ell_A^\mu}
    \biggl[\frac{v_A^\mu\,\ISP_1^{\powpropn_1}\cdots
                 \ISP_{\nisp}^{\powpropn_{\nisp}}}
    {\Den_1^{\powprop_1}\cdots
    \Den_{\ndenom}^{\powprop_{\ndenom}}}\biggr] =
    \frac{\frac{\partial}{\partial\ell_A^\mu} 
    \bigl[v_A^\mu\,\ISP_1^{\powpropn_1}\cdots
                 \ISP_{\nisp}^{\powpropn_{\nisp}}\bigr]}
    {\Den_1^{\powprop_1}\cdots
    \Den_{\ndenom}^{\powprop_{\ndenom}}}
    -\frac{\ISP_1^{\powpropn_1}\cdots
                 \ISP_{\nisp}^{\powpropn_{\nisp}}
                 \sum_{j=1}^c g_j}
{\Den_1^{\powprop_1}\cdots\Den_{\ndenom}^{\powprop_{\ndenom}}}\,.
\end{equation}
An IBP relation generated with $v$ does not have doubled 
propagators if none were present originally; and more 
generally, does not increase the powers of denominator 
factors.  Using all generating vectors,
we obtain a complete set of IBP relations, so that effectively
their use block-diagonalizes the system of equations.  The
\textsf{NeatIBP\/} code uses such generating vectors (in Baikov
variables) to obtain IBP relations.

\def\Rel{R}
\def\powvec{\pmb{\powprop}}
\def\powvecn{\pmb{\powpropn}}
The basic building block in our approach is an IBP equation
generated with a single generating vector out of the set of 
$\nvector$,
\begin{equation}
      0 = \Rel_i(\powvec;-\powvecn) = \int
      \prod_{j=1}^L \frac{\dd^D\ell_j}{(2\pi)^D} \;
    \frac{\partial}{\partial\ell_A^\mu}
    \biggl[\frac{v_{iA}^\mu\,\ISP_1^{\powpropn_1}\cdots
                 \ISP_{\nisp}^{\powpropn_{\nisp}}}
    {\Den_1^{\powprop_1}\cdots\Den_c^{\powprop_c}}\biggr]\,.
\end{equation}
We allow only equations with negative or zero values of the 
$\powpropn$ indices.
We define $d_i$ to be the degree of the $i^{\rm th}$ vector,
and $d_0$ the minimum degree of all vectors,
\begin{equation}
    \begin{aligned}
        d_i &= \cdeg v_i^\mu\,,\\
        d_0 &= \min_i d_i\,.
    \end{aligned}
\end{equation}
\def\Trilat#1{\mathbb{T}^{#1}}
We will make use of a triangular sublattice of the nonnegative integer lattice $\mathbb{Z}_{\ge0}^n$,
\begin{equation}
    \Trilat{n}[m] = \bigl\{\pmb{s}\in \mathbb{Z}^n_{\ge0}
    \big|\,|\pmb{s}|_1=m\bigr\}\,.
\end{equation}
Here, $|\pmb{s}|_1$ is the sum of the components of $\pmb{s}$.
We will call $m$ the \textit{level\/}, and the individual
elements \textit{shifts\/}.  For example, 
\begin{equation}
    \Trilat3[1]=\{(1,0,0)\,, (0,1,0)\,, (0,0,1)\}\,.
\end{equation}

\def\RSet{S}
\def\FRSet{S}
\def\CSet{C}
\def\level#1{{[#1]}}
\def\nmax{N_{\textrm{max}}}
Using such a sublattice, we can build sets of IBP relations
for a given vector at a given level,
\begin{equation}
    \RSet_i^\level{m}(\pmb{\powprop};-\pmb{\powpropn}) =
    \bigcup_{\pmb{s}\in\Trilat{\nisp}[m]}^{} 
    \Rel_i(\pmb{\powprop};\pmb{s}-\pmb{\powpropn})\,.
\end{equation}
We shift only the indices associated to the ISPs, and we exclude
equations where any such index is positive (that is,
where the ISP would appear in the denominator).  We then
build a set of relations from all $\nvector$ vectors
for the given integral topology.  We take all 
numerators to be of the same degree, 
\begin{equation}
    \FRSet^\level{m}(\pmb{\powprop};-\pmb{\powpropn}) = 
    \bigcup_{i=1}^{n_v} 
    s_0^m \RSet_i^\level{m+d_i-d_0}(\pmb{\powprop};\pmb{s}-\pmb{\powpropn})\,.
\end{equation}
We multiply by the power of $s_0$ to keep the dimensions
of equations of different levels the same.
For a given target integral, 
with generic or fixed
exponents $\pmb{\powprop}$ and $\pmb{\powpropn}$,
we assemble as many of these as required to find a linear
combination of the IBP relations so that the target integral
has a nonzero coefficient and that the linear combination
includes no top-level 
integral with any exponent $\powpropn_i$ larger
than that of the target integral.  (That is,
all top-level integrals with any ISP index more negative than 
in the target integral are unwanted, and we eliminate them.)
Solving the resulting
relation for the target integral gives us a \textit{lowering\/}
operator. Such lowering operators can be thought of as 
carefully chosen linear combinations of textbook lowering and 
raising operators~\cite{Smirnov:2006ry}.  We can write a 
complete set as follows,
\begin{equation}
    \CSet^{\level{\nmax}}(\pmb{\powprop};-\pmb{\powpropn}) = 
    \bigcup_{m=0}^{\nmax} 
    \FRSet^\level{m}(\pmb{\powprop};-\pmb{\powpropn})\,.
\end{equation}
Here, $\nmax$ is determined iteratively,
by increasing $m$ and adding sets of IBP relations until
the defining conditions can be met.  
The lowering operator we obtain lowers one or more indices,
and avoids raising any index in
top-level integrals.  In general, the resulting
lowering operator will not be unique, and we will be able to
impose further constraints to obtain different choices of
operator. This is a novel aspect of our approach.
Only the final result of repeatedly applying 
all relevant lowering operators to a given integral, yielding
an expression in terms of a chosen basis of integrals, will
be unique. Similar but less systematically organized sets were used in refs.~\cite{Kosower:2018obg,Abreu:2020lyk}.

\begin{figure}[htb]
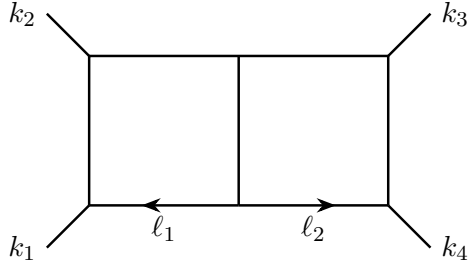

\centering
\DBox
\caption{Double box integral with massless external momenta}
\label{DoubleBoxFigure}
\end{figure}
 
\section{Massless Double Box}
\label{DoubleBoxSection}
In this section, we apply the strategy described in the 
previous section to the planar double-box integral.  This integral,
shown in fig.~\ref{DoubleBoxFigure},
has four external momenta ($k_i^2=0$) and seven denominators,
\begin{equation}
    \begin{aligned}
        \Den_1 &= \ell_1^2\,,\quad
        \Den_2 = (\ell_1-k_1)^2\,,\quad
        \Den_3 = (\ell_1-K_{12})^2\,,\quad
        \\ \Den_4 &= \ell_2^2\,,\quad
        \Den_5 = (\ell_2-k_4)^2\,,\quad
        \Den_6 = (\ell_2-K_{34})^2\,,\quad
        \Den_7 = (\ell_1+\ell_2)^2\,.
    \end{aligned}
\end{equation}
We denote a sum of momenta $K_{ij}=k_i+\cdots+k_j$.
It has two ISPs, which we choose to be,
\begin{equation}
    \ISP_1 = (\ell_1+k_4)^2\,,\quad 
    \ISP_2 = (\ell_2+k_1)^2\,.
\end{equation}
We choose $s_{12}$ as the standard invariant, and write
$s_{14}=\chi s_{12}$. We  use $\set{k_1^\mu, k_2^\mu,k_3^\mu}$ as the basis of external momenta and, 
\begin{equation}
V= \set{s_{12},k_1\cdot \ell_1,k_1\cdot \ell_2,k_2\cdot \ell_1,k_2\cdot \ell_2,k_3\cdot \ell_1,k_3\cdot \ell_2,\ell_1{}^2,\ell_1\cdot \ell_2,\ell_2^2}
\end{equation}
as our variables.

The integrand has two symmetries, corresponding to flips through
the vertical and horizontal axes,
\begin{equation}
    \begin{aligned}
        F_1:~& \ell_1\leftrightarrow \ell_2\,, 
        k_1\leftrightarrow k_4\,, k_2\leftrightarrow k_3\,;
        \\F_2:~& \ell_1\rightarrow K_{12}-\ell_1\,,
        \ell_2\rightarrow K_{34}-\ell_2\,,
         k_1\leftrightarrow k_2\,, k_3\leftrightarrow k_4\,.
    \end{aligned}
\end{equation}
These symmetries map the denominators and ISPs as follows,
\begin{equation}
    \begin{aligned}
        F_1:~& \Den_1\leftrightarrow \Den_4\,,
        \Den_2\leftrightarrow \Den_5\,,
        \Den_3\leftrightarrow\Den_6\,,
        \Den_7~\textrm{invariant}\,,
        \ISP_1\leftrightarrow\ISP_2\,;
        \\F_2:~& \Den_1\leftrightarrow\Den_3\,,
        \Den_4\leftrightarrow\Den_6\,,
        \Den_{2,5,7}~\textrm{and}~\ISP_{1,2}~\textrm{invariant}\,.
    \end{aligned}
\end{equation}
and accordingly map integrals as follows,
\begin{equation}
    \begin{aligned}
        F_1:~&\Integ[a_1,a_2,a_3,a_4,a_5,a_6,a_7,a_8,a_9]
    \longrightarrow\Integ[a_4,a_5,a_6,a_1,a_2,a_3,a_7,a_9,a_8]\,;
        \\F_2:~&\Integ[a_1,a_2,a_3,a_4,a_5,a_6,a_7,a_8,a_9]
    \longrightarrow\Integ[a_3,a_2,a_1,a_6,a_5,a_4,a_7,a_8,a_9]\,.
    \end{aligned}
\end{equation}

There are three independent generating vectors, one of degree
($\cdeg$) one, and two of degree two.  The first vector is, 
\def\splitl{\\&\hphantom{=!}}
\def\tm{\!-\!}
\def\tp{\!+\!}
\begin{equation}
\hspace*{-3mm}\begin{aligned}
v_{1;1}^\mu &= 2k_1\cdot \ell_1 \,k_2^\mu\tp (2\,k_1\tm \ell_1)\cdot \ell_1 \,k_3^\mu\tm (2\,(k_2\tp k_3)\cdot \ell_1\tm \ell_1^2) \,k_1^\mu\tp (s_{12}\tm 2 (k_1\tm k_3)\cdot \ell_1) \,\ell_1^\mu
\,,\\ 
v_{1;2}^\mu &= 2k_1\cdot \ell_2 \,k_2^\mu\tm (2\,(k_2\tp k_3)\cdot \ell_2\tp \ell_2^2) \,k_1^\mu\tp (s_{12}\tp 2 (k_1\tm k_3)\cdot \ell_2) \,\ell_2^\mu\tp (2\,k_1\tp \ell_2)\cdot \ell_2 \,k_3^\mu\,.
\end{aligned}
\end{equation}
The first subscript is the vector number, and the second
the loop index.  The expressions for the other two vectors are
lengthier.  All vectors are given in the auxiliary file
\textsf{db\nobreakdash-vectors.m}.  We computed them using 
\textsf{Singular\/}~\cite{Singular}.

\subsection{Bulk Lowering Operators}
\label{DBBulkSection}

Our first task is to obtain a lowering operator for ordinary
integrals, with denominator exponents all taking the value $1$
and generic numerator exponents $\powpropn_1$ and 
$\powpropn_2$.
Thanks to the $F_1$ symmetry, we can 
take $\powpropn_1\ge\powpropn_2$ without loss of generality.
Let's start by writing down the set of equations with level-shift
zero,
\begin{equation}
\begin{aligned}
    \FRSet^\level0(&\pmb{1};-\powpropn_1,-\powpropn_2) =\\
    \bigl\{& \Rel_1(\pmb{1};-\powpropn_1,-\powpropn_2)\,,
             \Rel_2(\pmb{1};1-\powpropn_1,-\powpropn_2)\,,
             \Rel_2(\pmb{1};-\powpropn_1,1-\powpropn_2)\,,
            \\& \Rel_3(\pmb{1};1-\powpropn_1,-\powpropn_2)\,,
             \Rel_3(\pmb{1};-\powpropn_1,1-\powpropn_2)\bigr\}\,,
\end{aligned}
    \label{Level0Relations}
\end{equation}
and forming a general linear combination,
\def\coeff#1{c_{#1}}
\begin{equation}
\begin{aligned}
    &\coeff1 \Rel_1(\pmb{1};-\powpropn_1,-\powpropn_2)
    +\coeff2 \Rel_2(\pmb{1};1-\powpropn_1,-\powpropn_2)
    +\coeff3\Rel_2(\pmb{1};-\powpropn_1,1-\powpropn_2) 
    \\&+\coeff4 \Rel_3(\pmb{1};1-\powpropn_1,-\powpropn_2)
    +\coeff5 \Rel_3(\pmb{1};-\powpropn_1,1-\powpropn_2) 
    =
    \\& 
-(\powpropn_1\,c_{1}-2\,\eps\,c_{1}+2\,c_{3}
-2\,\powpropn_2\,c_{3}+4\,\eps\,c_{3})\,
\Integ[\pmb{1};-1-\powpropn_1,-\powpropn_2]
\\&
+2\,(1+\powpropn_2-2\,\eps)\,c_{4}\,
\Integ[\pmb{1};1-\powpropn_1,-2-\powpropn_2]
-2\,(1-\powpropn_2+\eps)\,c_{3}\,s_{12}\,
\Integ[\pmb{1};-1-\powpropn_1,1-\powpropn_2]
\\&
+(\powpropn_2-2\,\eps)\,(c_{1}+2\,c_{2}+2\,c_{5})\,
\Integ[\pmb{1};-\powpropn_1,-1-\powpropn_2]
\\&
-2\,\chi\,(\powpropn_2-\eps)\,(c_{2}+c_{4})\,s_{12}^{2}\,
\Integ[\pmb{1};1-\powpropn_1,-\powpropn_2]
\\&
+2\,(\powpropn_2\,c_{2}-2\,\eps\,c_{2}
-c_{4}+\powpropn_1\,c_{4}
+\powpropn_2\,c_{4}-2\,\chi\,c_{4}
\\&\hphantom{+2(}+\powpropn_1\,\chi\,c_{4}-\powpropn_2\,\chi\,c_{4}
-3\,\eps\,c_{4})\,s_{12}\,
\Integ[\pmb{1};1-\powpropn_1,-1-\powpropn_2]
\\&
+(\powpropn_1\,\chi\,c_{1}-\powpropn_2\,\chi\,c_{1}+2\,\powpropn_2\,c_{2}-2\,\eps\,c_{2}
-2\,c_{3}+2\,\powpropn_2\,c_{3}-4\,\eps\,c_{3}-2\,c_{5}
+2\,\powpropn_1\,c_{5}+2\,\powpropn_2\,c_{5}
\\&
\hphantom{+(}
+2\,\powpropn_1\,\chi\,c_{5}-2\,\powpropn_2\,\chi\,c_{5}
-6\,\eps\,c_{5})\,s_{12}\,
\Integ[\pmb{1};-\powpropn_1,-\powpropn_2]
\\&
+2\,\chi\,(1-\powpropn_2+\eps)\,(c_{3}+c_{5})\,s_{12}^{2}\,
\Integ[\pmb{1};-\powpropn_1,1-\powpropn_2]
+\textrm{daughters}\,,
\end{aligned}
\end{equation}
where as usual $\eps=(D-4)/2$.
We refer to integrals with fewer denominator factors as `daughters'.
In this equation, there are five integrals with indices more
negative than the target 
$\Integ[\pmb{1};-\powpropn_1,-\powpropn_2]$,
\begin{equation}
    \begin{aligned}
&\Integ[\pmb{1};-1-\powpropn_1,1-\powpropn_2]
\,, \Integ[\pmb{1};-1-\powpropn_1,-\powpropn_2]
\,, \Integ[\pmb{1};1-\powpropn_1,-2-\powpropn_2]
\,,     \\
&\Integ[\pmb{1};1-\powpropn_1,-1-\powpropn_2]
\,, \Integ[\pmb{1};-\powpropn_1,-1-\powpropn_2]\,.
    \end{aligned}
\label{Unwanted1}
\end{equation}
We must choose the $c_i$ so that the coefficients of these
integrals vanish.  The only solution has all $c_i$ vanishing,
so that these equations do not suffice.  Next, we add in
the eight equations at level-shift $1$,
\begin{equation}
    \begin{aligned}
    \FRSet^\level1(&\pmb{1};-\powpropn_1,-\powpropn_2) =
    \\
    \bigl\{& \Rel_1(\pmb{1};1-\powpropn_1,-\powpropn_2)\,,
    \Rel_1(\pmb{1};-\powpropn_1,1-\powpropn_2)\,,
            \\& 
             \Rel_2(\pmb{1};2-\powpropn_1,-\powpropn_2)\,,
             \Rel_2(\pmb{1};1-\powpropn_1,1-\powpropn_2)\,,
             \Rel_2(\pmb{1};-\powpropn_1,2-\powpropn_2)\,,
            \\& 
            \Rel_3(\pmb{1};2-\powpropn_1,-\powpropn_2)\,,
            \Rel_3(\pmb{1};1-\powpropn_1,1-\powpropn_2)\,,
             \Rel_3(\pmb{1};-\powpropn_1,2-\powpropn_2)\bigr\}\,.
    \end{aligned}
    \label{Level1Relations}
\end{equation}
That is, we assemble 
$\CSet^{\level1}(\pmb{1};-\powpropn_1,-\powpropn_2)$.
Coefficients $c_{1\cdots5}$ are associated to the relations in
\eqn{Level0Relations}, 
and coefficient $c_{6\cdots13}$ to those in \eqn{Level1Relations}.
Now, there are eight unwanted integrals, those in 
\eqn{Unwanted1} along with,
\begin{equation}
\begin{aligned}
    &\Integ[\pmb{1};-1-\powpropn_1,2-\powpropn_2]
\,, \Integ[\pmb{1};2-\powpropn_1,-2-\powpropn_2]
\,, \Integ[\pmb{1};2-\powpropn_1,-1-\powpropn_2]\,.
\end{aligned}
\label{Unwanted2}
\end{equation}

Here we can find a nontrivial solution for the $c_i$
which eliminates all integrals in \eqns{Unwanted1}{Unwanted2}
while retaining $\Integ[\pmb{1};-\powpropn_1,-\powpropn_2]$, 
\begin{equation}
\begin{aligned}
c_{1} &= -2 \frac{(1-\powpropn_2+2\,\eps)\,c_{3}}{\powpropn_1-2\,\eps}\,,\\
c_{4} &= 0\,,\\
c_{5} &= -c_{2}+\frac{(1-\powpropn_2+2\,\eps)\,c_{3}}{\powpropn_1-2\,\eps}\,,\\
c_{7} &= -2 \frac{(1-\powpropn_2+\eps)\,c_{3}}{\powpropn_1-2\,\eps}\,,\\
c_{8} &= 0\,,\\
c_{9} &= 0\,,\\
c_{11} &= 0\,,\\
c_{13} &= -c_{2}-\frac{c_{6}}{2}\,.
\end{aligned}
\label{BulkGeneralSolution}
\end{equation}
We then solve the IBP relation to obtain a lowering
operator.
We still have five unfixed coefficients 
($c_2, c_3, c_6, c_{10}, c_{12}$),
so that we can obtain a four-parameter family of 
lowering operators.
(One coefficient will ultimately just give an overall scale of
the equation and will cancel out of the lowering operator.)
The lowering operator has the form,
\begin{equation}
    \begin{aligned}
\Lowering{[g]}(\powpropn_1,\powpropn_2) =\,& 
\Integ[\pmb{1};-\powpropn_1,-\powpropn_2]
 \longrightarrow \,\\
 & \gamma_{1}\,s_{12}\,\Integ[\pmb{1};1\tm \powpropn_1,\tm \,\powpropn_2]
\tp \gamma_{2}\,s_{12}\,\Integ[\pmb{1};\tm \,\powpropn_1,1\tm \powpropn_2]
\tp \gamma_{3}\,s_{12}^{2}\,\Integ[\pmb{1};1\tm \powpropn_1,1\tm \powpropn_2]
\\& \tp \gamma_{4}\,s_{12}^{2}\,\Integ[\pmb{1};\tm \,\powpropn_1,2\tm \powpropn_2]
+\textrm{daughters}\,,
    \end{aligned}
    \label{BulkLoweringOperatorForm}
\end{equation}
in which,
\begin{equation}
    \begin{aligned}
\gamma_{1} &= -\xi_{1}^{\tm 1}(2\,(2\tm \powpropn_1\tm \powpropn_2\tp \chi\tm \powpropn_1\,\chi\tp 3\,\eps\tp \chi\,\eps)\,c_{2}\tp (2\tm \powpropn_1\tm \powpropn_2\tp 3\,\eps)\,c_{6}\tm 2\,(1\tm \powpropn_2\tp 2\,\eps)\,c_{10})\,,\hspace*{-4mm}\\
\gamma_{2} &= \xi_{1}^{\tm 1}((1\tm \powpropn_2\tp \eps)\,(2\,\chi\,c_{2}\tp 2\,
c_{10})\tp 2\,(2\tm \powpropn_1\tm \powpropn_2\tm \chi\tm \powpropn_1\,\chi\tp \powpropn_2\,\chi\tp 3\,\eps)\,c_{12})\,,\\
\gamma_{3} &= \xi_{1}^{\tm 1}((1\tm \powpropn_2\tp \eps)\,(2\,\chi\,c_{2}\tp \chi\,c_{6})\tm 2\,\chi\,(1\tm \powpropn_2\tp \eps)\,c_{10})\,,\\
\gamma_{4} &= - 2\xi_{1}^{\tm 1}\chi\,(2\tm \powpropn_2\tp \eps)\,c_{12}\,,
    \end{aligned}
\end{equation}
and,
\begin{equation}
    \begin{aligned}
\xi_{1} &= 2\,(1\tm \powpropn_1\tm \powpropn_1\,\chi\tp \powpropn_2\,\chi\tp 2\,
\eps)\,c_{2}\tp (1\tm \powpropn_1\tp 2\,\eps)\,c_{6}\tm 2\,(1\tm \powpropn_2\tp
2\,\eps)\,(c_{10}\tp c_{12})\,.
    \end{aligned}
\end{equation}
All the operators in this and following subsections are
given in an auxiliary file 
\textsf{db\nobreakdash-operators.m\/}, including
daughter integral terms omitted above. 

We can specialize this operator by imposing additional 
conditions. For example, we could avoid increasing the 
power of the top-level ISPs in any daughter integral; 
or we could eliminate all integrals
with a $2-\powpropn_i$ index, which would allow the operator
to be used for values of $\powpropn_i$ down to $1$ (and thereby
reduce the number of different boundary and edge operators we
will need to consider in following subsections).  Indeed,
we can combine these two restrictions, which will yield 
an operator with no free coefficients, given by the form
in \eqn{BulkLoweringOperatorForm} with, 
\begin{equation}
    \begin{aligned}
\gamma_{1} &= - \frac{2-\powpropn_1-\powpropn_2+3\,\eps}
{1-\powpropn_1+2\,\eps}\,,\\
\gamma_{2} &= 0\,,\\
\gamma_{3} &= \frac{\chi\,(1-\powpropn_2+\eps)}
{1-\powpropn_1+2\,\eps}\,,\\
\gamma_{4} &= 0
\,.
\end{aligned}
    \label{RefinedBulkLoweringOperatorCoefficients}
\end{equation}
The operator can be applied to simplify all integrals
with $\powpropn_1\ge\powpropn_2\ge 1$.  
It is similar to the operator for the one-mass
double box given in rule form
by Smith and Zeng \cite{Smith:2025xes}.
We will need
additional `boundary' operators for the case 
$(\powpropn_1,\powpropn_2)=(\powpropn_1,0)$, which we will
construct in the next subsection.

Alternatively, we can construct
a bulk lowering operator that lowers only the second ISP 
index in top-topology integrals.
An example of such a `wall-hugging' operator has the 
form~\eqref{BulkLoweringOperatorForm} with,
\begin{equation}
    \begin{aligned}
\gamma_{1} &= 0\,,\\
\gamma_{2} &= - \frac{2-\powpropn_2\,(1-\chi)-\chi
-\powpropn_1\,(1+\chi)+3\,\eps}
{1-\powpropn_2+2\,\eps}\,,\\
\gamma_{3} &= 0\,,\\
\gamma_{4} &= \frac{\chi\,(2-\powpropn_2+\eps)}
{1-\powpropn_2+2\,\eps}
\,.
    \end{aligned}
\label{BulkWallLoweringOperatorCoefficients}
\end{equation}
This operator requires $\powpropn_1\ge\powpropn_2\ge 2$;
for $\powpropn_2=1$, one can either use the operator above
with coefficients given 
in \eqn{RefinedBulkLoweringOperatorCoefficients}, or
obtain an additional boundary operator.
We leave the question of an optimal choice of the coefficients $c_i$ to future investigation.

\subsection{Boundary Lowering Operators}
\label{DBBoundarySection}

As explained in the previous subsection,
the lowering operator~\eqref{BulkLoweringOperatorForm} with 
the coefficients $\gamma_{1,2,3,4}$ given
in \eqn{RefinedBulkLoweringOperatorCoefficients} can be 
applied to all integrals with 
$\powpropn_1\ge\powpropn_2\ge 1$, but not with 
$\powpropn_2 = 0$.   
We need a specialized `boundary' operator to handle
integrals with ISP exponents $(\powpropn_1,0)$.  
Here, level shifts $0$ and $1$ do not suffice to find
a lowering operator.  When we add level shift $2$,
we obtain nine equations from 
$\CSet^{\level2}(\pmb 1; -\powpropn_1,0)$ and eight constraints, and thus
a unique lowering operator. We obtain
an operator of the form,
\begin{equation}
    \begin{aligned}
    \hspace*{-3mm}\Lowering{[\partial]}(\powpropn_1,0) = 
\Integ[\pmb{1};-\powpropn_1,0]
 &\longrightarrow 
 \gamma_{5}
\,s_{12}\,\Integ[\pmb{1};1-\powpropn_1,0]
+\gamma_{6}\,s_{12}^{2}\,\Integ[\pmb{1};2-\powpropn_1,0]
+\textrm{daughters}\,,
    \end{aligned}
    \label{BoundaryLoweringOperatorForm}
\end{equation}
with,
\begin{equation}
    \begin{aligned}
\gamma_{5} &= - \frac{2-\powpropn_1\,(1-\chi)-\chi+3\,\eps}
{1-\powpropn_1+2\,\eps}\,,\\
\gamma_{6} &= \frac{\chi\,(2-\powpropn_1+\eps)}
{1-\powpropn_1+2\,\eps}
\,.
    \end{aligned}
    \label{BoundaryLoweringOperatorCoefficients}
\end{equation}
The daughter terms contain indices as large 
as $3-\powpropn_1$, and so this operator
can be applied freely to all integrals with ISP exponents 
$(\powpropn_1\ge 3,0)$.  We do not
expect to find an operator that can reduce the integral with 
ISP exponents $(1,0)$, as
it is a master in the usual Laporta ordering.
It turns out that the coefficients of integrals containing
an index $3-\powpropn_1$ are all proportional to 
$2-\powpropn_1$, so that we can apply the operator 
(with appropriate care) to the case $(2,0)$ as well.
In general, we might have to obtain special `edge' operators,
where all $\powpropn_i$ have fixed values, 
for such cases. The operator 
in \eqn{BoundaryLoweringOperatorForm} goes beyond those presented in Smith and Zeng~\cite{Smith:2025xes} where instead they mix in the use of a standard Laporta approach. 
\subsection{Propagator Lowering Operators}
\label{DBPropagatorSection}

The use of generating vectors ensures that IBP equations 
don't introduce propagators with powers above $1$
if they weren't present initially.  In general, because of
gauge invariance, expressions for amplitudes in Yang--Mills
theory and gravity do not require such integrals.  Nonetheless, 
they may arise through the process of expansions (such as the 
post-Newtonian expansion in gravity), through differentiation 
with respect to external momenta in deriving 
differential equations, or after integrating out inner two- 
or three-point loops.
While generating vectors do not introduce such dotted 
integrals, they can of course be used to remove them.  In 
this subsection, we discuss the construction of lowering
operators that lower propagator indices greater than $1$.

Let's begin by finding an operator which lowers the first
propagator's index, taking 
$\Integ[1+\powprop_1,\pmb{1};-\pmb{\powpropn}]$ as our target.
The level-0 set of equations does not suffice, but
using $\CSet^{\level1}(1+\powprop_1,\pmb{1};-\pmb{\powpropn})$, 
we end up with five unfixed coefficients 
($c_2, c_3, c_6, c_{10}, c_{12}$)
 and hence a four-parameter family of 
lowering operators, 
\begin{equation}
    \begin{aligned}
\Lowering{[q]}&(1+\powprop_1;\pmb{1};\powpropn_1,\powpropn_2) = 
\\&\hspace*{-2mm}\Integ[1+\powprop_1;\pmb{1};
-\powpropn_1,-\powpropn_2]
 \longrightarrow \\
& \gamma_{7}\,s_{12}\,
\Integ[1\tp \powprop_1,\pmb{1};\tm\powpropn_1,1\tm \powpropn_2]
\tp \gamma_{8}\,s_{12}^{2}\,
\Integ[1\tp \powprop_1,\pmb{1};\tm\powpropn_1,
2\tm \powpropn_2]
\\& 
\tp \gamma_{9}\,s_{12}\,
\Integ[1\tp \powprop_1,\pmb{1};1\tm \powpropn_1,\tm\powpropn_2]
\tp \gamma_{10}\,s_{12}^{2}\,
\Integ[1\tp \powprop_1,\pmb{1};1\tm \powpropn_1,
1\tm \powpropn_2]
\\& 
\tp \gamma_{11}\,s_{12}^{\tm 1}\,
\Integ[\powprop_1,\pmb{1};\tm 1\tm \powpropn_1,
1\tm \powpropn_2]
\tp \gamma_{12}\,s_{12}^{\tm }\,
\Integ[\powprop_1,\pmb{1};\tm\powpropn_1,\tm\powpropn_2]
\tp \gamma_{13}\,
\Integ[\powprop_1,\pmb{1};\tm\powpropn_1,1\tm \powpropn_2]
\\& 
\tp \gamma_{14}\,s_{12}\,
\Integ[\powprop_1,\pmb{1};\tm\powpropn_1,2\tm \powpropn_2]
\tp \gamma_{15}\,
\Integ[\powprop_1,\pmb{1};1\tm \powpropn_1,\tm\powpropn_2]
\tp \gamma_{16}\,s_{12}\,
\Integ[\powprop_1,\pmb{1};1\tm \powpropn_1,1\tm \powpropn_2]
\\&
+\textrm{simpler}
+\textrm{daughters}\,,
\hspace*{-10mm}
    \end{aligned}
\end{equation}
where `simpler' refers to integrals in which the first
index is two or more units lower than the value in the target.
The coefficients $\gamma_{7\cdots16}$ are given by, 
\begin{equation}
    \begin{aligned}
\gamma_{7} &= \tm \xi_{2}^{\tm 1}(2\,\powprop_1\,(5\,\powprop_1\tm \powpropn_1\tm 5\,\chi\tm \powprop_1\,\chi\tm 3\,\powpropn_1\,\chi\tp 5\,\powpropn_2\,\chi\tp \eps\tm 3\,\chi\,\eps)\,c_{2} \\&\hphantom{=!}
\tm(\powprop_1\tm \powpropn_1\tp 2\,\eps)\,(2\,(\powprop_1\tm \chi\tp \powpropn_2\,\chi\tm \chi\,\eps)\,c_{1}\tm 2\,(1\tp 4\,\powprop_1\tm \powpropn_2\tp \eps)\,c_{4})
\\&\hphantom{=!}
\tp 2\,(\powprop_1\tm \powpropn_1\tp 2\,\eps)\,(2\tp \powprop_1\tm \powpropn_1\tm \powpropn_2\tm \chi\tm \powpropn_1\,\chi\tp \powpropn_2\,\chi\tp 3\,\eps)\,c_{5})\,,\\
\gamma_{8} &= \tm 2\xi_{2}^{\tm 1}(\powprop_1\tm \powpropn_1\tp 2\,\eps)\,(\powprop_1\tm 2\,\chi\tp \powpropn_2\,\chi\tm \chi\,\eps)\,c_{5}\,,\\
\gamma_{9} &= \tm \xi_{2}^{\tm 1}(\tm (\powprop_1\tm \powpropn_1\tp 2\,\eps)\,(2\,(2\tm \powpropn_1\tm \powpropn_2\tp \chi\tp \powprop_1\,\chi\tm \powpropn_1\,\chi\tp 3\,\eps\tp \chi\,\eps)\,c_{1}
\\&\hphantom{=!++}
\tp (2\tm \powpropn_1\tm \powpropn_2\tp 3\,\eps)\,c_{3})
\tp 2\,(\powprop_1\tm \powpropn_1\tp 2\,\eps)\,(1\tm \powpropn_2\tp 2\,\eps)\,c_{4})\,,\hspace*{-10mm}\\
\gamma_{10} &= \tm \xi_{2}^{\tm 1}(\tm (\powprop_1\tm \powpropn_1\tp 2\,\eps)\,(2\,(\powprop_1\tm \chi\tp \powpropn_2\,\chi\tm \chi\,\eps)\,c_{1}\tp (\powprop_1\tm \chi\tp \powpropn_2\,\chi\tm \chi\,\eps)\,c_{3})
\\&\hphantom{=!}
\tp 2\,(\powprop_1\tm \powpropn_1\tp 2\,\eps)\,(\powprop_1\tm \chi\tm \powprop_1\,\chi\tp \powpropn_2\,\chi\tm \chi\,\eps)\,c_{4})\,,\\
\gamma_{11} &= \tm 10\xi_{2}^{\tm 1}(1\tm \powprop_1\tp \powpropn_1\tm 2\,\eps)\,(\powprop_1\tm \powpropn_1\tp 2\,\eps)\,c_{2}\,,\\
\gamma_{12} &= 8\xi_{2}^{\tm 1}(\powprop_1\tm \powpropn_1\tp 2\,\eps)^{2}\,c_{1}\,,\\
\gamma_{13} &= \tm \xi_{2}^{\tm 1}(2\,\powprop_1\tm \powpropn_1\tp 2\,\eps^{2}\,c_{1}\tm 2\,(\powprop_1\tm \powpropn_1\tp 2\,\eps)\,(1\tp 4\,\powprop_1\tm \powpropn_1\tm \powprop_1\,\chi\tp 2\,\powpropn_1\,\chi\tp \eps\tm 3\,\chi\,\eps)\,c_{2} 
\hspace*{-10mm}
\\&\hphantom{=!}
\tm 10\,\powprop_1\tm \powpropn_1\tp 2\,\eps^{2}\,c_{4}\tm 2\,\powprop_1\tm\powpropn_1\tp 2\,\eps^{2}\,c_{5})\,,\\
\gamma_{14} &= 2\xi_{2}^{\tm 1}(\powprop_1\tm \powpropn_1\tp 2\,\eps)^{2}\,c_{5}\,,\\
\gamma_{15} &= 2\xi_{2}^{\tm 1}(\powprop_1\tm \powpropn_1\tp 2\,\eps)\,(1\tm \powprop_1\tm 2\,\chi\tm \powprop_1\,\chi\tp 2\,\powpropn_1\,\chi\tm 3\,\chi\,\eps)\,c_{1}\,,\\
\gamma_{16} &= \xi_{2}^{\tm 1}(\tm (\powprop_1\tm \powpropn_1\tp 2\,\eps)\,(2\,(1\tp \powprop_1\tm \powpropn_1\tp 2\,\eps)\,c_{1}\tp (1\tp \powprop_1\tm \powpropn_1\tp 2\,\eps)\,c_{3})
\\&\hphantom{=!}
\tp 2\,(\powprop_1\tm \powpropn_1\tp 2\,\eps)\,(2\tm \powpropn_1\tm 2\,\chi\tm \powprop_1\,\chi\tp 2\,\powpropn_1\,\chi\tp 2\,\eps\tm 3\,\chi\,\eps)\,c_{4})\,,\\
    \end{aligned}
\label{PropagatorOperatorCoefficents}
\end{equation}
in which,
\begin{equation}
    \begin{aligned}
\xi_{2} &= \powprop_1^{2}\,(6\,c_{1}\tm c_{3})\tm \powprop_1\,
(2\,(1\tp \powpropn_1\,(2\tm \chi)\tp \powpropn_2\,\chi
\tm 4\,\eps)\,c_{1}\tm 10\,(1\tm \powpropn_2
\tp 2\,\eps)\,c_{2}
\\&\hphantom{=!}
\tp c_{3}\tm 2\,\powpropn_1\,c_{3}
\tp 4\,\eps\,c_{3}\tm 2\,c_{4}\tp 2\,\powpropn_2\,c_{4}
\tm 4\,\eps\,c_{4}\tm 2\,c_{5}\tp 2\,\powpropn_2\,c_{5}
\tm 4\,\eps\,c_{5})
\\&\hphantom{=!}
\tp (\powpropn_1\tm 2\,\eps)\,
(2\,(1\tm \powpropn_1\tm \powpropn_1\,\chi
\tp \powpropn_2\,\chi\tp 2\,\eps)\,c_{1}\tp 
(1\tm \powpropn_1\tp 2\,\eps)\,c_{3}
\\&\hphantom{=!}
\tm 2\,
(1\tm \powpropn_2\tp 2\,\eps)\,(c_{4}\tp c_{5}))
    \end{aligned}
\end{equation}
In this general form, we have relaxed the definition of
a lowering operator, allowing increased ISP indices in terms
that lower the first propagator index.

  We can specialize this operator, for example,
by eliminating all top-level integrals on the right-hand side
which still have the original first-propagator power
$1+\powprop_1$, that is requiring,
\begin{equation}
    \gamma_{7}=\gamma_{8}=\gamma_{9}=\gamma_{10}=0\,.
\end{equation}
We can implement this by choosing,
\begin{equation}
    \begin{aligned}
 c_{2} &= \xi_{3}^{\tm 1}\frac{(\powprop_1\tm \powpropn_1\tp 2\,\eps)\,(\powprop_1\tm \chi\,(1\tm \powpropn_2\tp \eps))\,(1\tm \powpropn_1\,(1\tp 3\,\chi)\tp \eps\tp \chi\,(3\tp 4\,\powprop_1\tp 2\,\eps))\,c_{1}}{\powpropn_1\tm \powprop_1\,(5\tm \chi)\tp 5\,\chi\tp 3\,\powpropn_1\,\chi\tm 5\,\powpropn_2\,\chi\tm \eps\tp 3\,\chi\,\eps}\,,\\
c_{3} &= 2\xi_{3}^{\tm 1}(\powprop_1^{2}\,(1\tm \chi)\,\chi\tm \chi\,(1\tp \chi)\,(1\tm \powpropn_1\tp \eps)\,(1\tm \powpropn_2\tp \eps)\tp \powprop_1\,(1\tm \powpropn_1\,(1\tm \chi^{2})\tp \eps
\\&\hphantom{=!}
\tm \chi\,(1\tm \powpropn_2\tp 2\,\eps)\tm \chi^{2}\,(2\tm \powpropn_2\tp 2\,\eps)))\,c_{1}\,,\\
c_{4} &= \xi_{3}^{\tm 1}\chi\,(1\tp \powprop_1\tm \powpropn_1\tp \eps)\,(\powprop_1\tm \chi\,(1\tm \powpropn_2\tp \eps))\,c_{1}\,,\\
c_{5} &= 0\,,\\
\xi_{3} &= \chi\,(1\tm \powpropn_1\tp \eps)\,(1\tm \powpropn_2\tp \eps)\tm \powprop_1\,(1\tm \powpropn_1\,(1\tm \chi)\tp \eps\tm \chi\,(2\tm \powpropn_2\tp 3\,\eps))\,,
    \end{aligned}
\end{equation}
for the coefficients $c_i$ in \eqn{PropagatorOperatorCoefficents}.

One could proceed in this fashion, obtaining required boundary
operators, finding operators that
lower other propagator indices, and then pairs of them, 
etc.; but as in principle we will need to handle the case
where all propagator indices are higher than $1$, we may
as well proceed directly to that case.  We take 
$\Integ[\pmb{1}+\pmb{\powprop};-\pmb{\powpropn}]$ as our target.
With level-0 and -1 shifts, we obtain a four-parameter 
solution.  We again choose
to further remove all integrals with the original powers 
and lower ISP indices, ensuring
that every term on the right-hand side lowers at least 
one propagator index.  The resulting operator 
$\Lowering{[p]}(1+\pmb{\powprop};\pmb{\powpropn})$
is too lengthy to display in print (it is given in
the auxiliary
file \textsf{db-operators.m\/}).  It contains ISP indices 
as large as $2-\powpropn_1$
and $2-\powpropn_2$, but the coefficients can be chosen that it 
can be applied for
$\powpropn_1 \ge \powpropn_2>0$.  (The lowering operator 
does require at least one
$\powprop_i$ to be larger than $0$.  It has three remaining independent
parameters $c_i$.)   This leaves two 
required boundary operators
(edge operators as far as the ISP indices alone 
are concerned), for
$(-\powpropn_1,-\powpropn_2)=(-1,0), (0,0)$.  We can obtain 
these operators ($\Lowering{[p1]}(\pmb{1}+\pmb{\powprop};1,0)$
and $\Lowering{[p0]}(\pmb{1}+\pmb{\powprop};0,0)$) by
considering the set of equations 
$\CSet^{\level1}(\pmb{1}+\pmb{\powprop};-1,-1)$.

\section{Massless Pentabox}
\label{PentaboxSection}
\begin{figure}[htb]
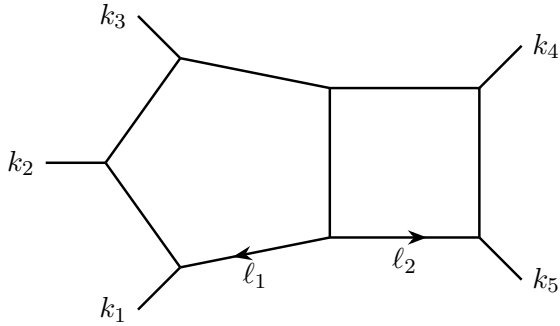

\centering
\PentaBox
\caption{Pentabox graph with outgoing massless external momenta.}
\label{pentaboxgraph}
\end{figure}

In this section, we look at the bulk and boundary lowering operators needed for the
massless planar pentabox, shown in fig.~\ref{pentaboxgraph}.  It has five external momenta ($k_i^2=0$), 
eight denominators,
\begin{equation}
    \begin{aligned}
        \Den_1 &= \ell_1^2\,,\quad
        \Den_2 = (\ell_1-k_1)^2\,,\quad
        \Den_3 = (\ell_1-K_{12})^2\,,\quad
        \Den_4 = (\ell_1-K_{123})^2\,,\quad
        \\ \Den_5 &= \ell_2^2\,,\quad
        \Den_6 = (\ell_2-k_5)^2\,,\quad
        \Den_7 = (\ell_2-K_{45})^2\,,\quad
        \Den_8 = (\ell_1+\ell_2)^2\,,
    \end{aligned}
\end{equation}
and three ISPs, which we choose to be,
\begin{equation}
    \ISP_1 = (\ell_1+k_5)^2\,,\quad 
    \ISP_2 = (\ell_2+k_1)^2\,,\quad
    \ISP_3 = (\ell_2+K_{12})^2\,.
\end{equation}

In order to lighten the computational load, we work at a specific kinematic point, with
\begin{equation}
\chi_{15} = -\frac{1}{223}\,,\quad \chi_{23} = -\frac{1}{2}\,,\quad 
\chi_{34} = -\frac{1}{73}\,,\quad \chi_{45} = -\frac{1}{7}\,.
\end{equation}
We find 11 IBP-generating vectors, six of coefficient degree two, and five of coefficient degree three.
The vectors are given in the auxiliary file \textsf{pb-vectors.m\/}.
The standard choice of top-level master integrals is,
\begin{equation}
    \bigl\{\Integ[\pmb{1};-1,0,0]\,, \,\,\Integ[\pmb{1};0,0,-1]\,, \,\,\Integ[\pmb{1};0,0,0]\bigr\}\,.  
\end{equation}

We can find an equation for a bulk lowering operator
for undotted integrals using the set 
$\CSet^{\level1}(\pmb{1};-\pmb{\powpropn})$.  This uses
69 IBP relations, and we find a five-parameter family
of lowering operators.  It does place some constraints
on the $\powpropn_i$: $\powpropn_1\neq\powpropn_2+\powpropn_3$ 
and $\powpropn_2\neq\powpropn_3$. 
We can avoid this by including an additional level, that is 
using $\CSet^{\level2}(\pmb{1};-\pmb{\powpropn})$.  We then
obtain a 39-parameter family of
lowering operators valid for 
$(\powpropn_1>0,\powpropn_2>0,\powpropn_3>0)$.  We
need a first set of boundary operators, 
\begin{equation}
    \Lowering{[\partial]}(0,\powpropn_2,\powpropn_3)\,,
    \Lowering{[\partial]}(\powpropn_1,0,\powpropn_3)\,,
    \Lowering{[\partial]}(\powpropn_1,\powpropn_2,0)\,,
\end{equation}
which we can obtain without restrictions from 
$\CSet^{\level2}(\pmb{1};0,-\powpropn_2,-\powpropn_3)$,
$\CSet^{\level2}(\pmb{1};-\powpropn_1,0,-\powpropn_3)$, and
$\CSet^{\level2}(\pmb{1};-\powpropn_1,-\powpropn_2,0)$ respectively.
Finally we need another three operators for their boundaries (i.e. boundaries-of-boundaries),
\begin{equation}
    \Lowering{[\partial\partial]}(0,0,\powpropn_3)\,,
    \Lowering{[\partial\partial]}(0,\powpropn_2,0)\,,
    \Lowering{[\partial\partial]}(\powpropn_1,0,0)\,,
\end{equation}
which we can obtain from
$\CSet^{\level4}(\pmb{1};0,0,-\powpropn_3)$,
$\CSet^{\level4}(\pmb{1};0,-\powpropn_2,0)$, and
$\CSet^{\level3}(\pmb{1};-\powpropn_1,0,0)$ respectively.
We have used the 
\textsf{FiniteFlow\/} package~\cite{Peraro:2019svx}
for some calculations in finding these operators.
We leave an exploration of the propagator-lowering operators to future work.

\section{Implementation}
\label{ImplementationSection}

Within \textsl{Mathematica\/} or other symbolic math systems, 
one can implement the lowering operators as rules relying on
pattern matching.  For numerical implementation in other 
frameworks, such as \textsf{Python\/} or \textsf{C++\/}, we
need a different paradigm.  There are two that appear 
suitable.   These are a recursive implementation, with
caching of intermediate results (for example, via a
hash table); and use of sparse matrices.

Each lowering operator for specific values of the 
indices can be represented as a sparse matrix.  
Nonzero entries are indexed by a pair of vectors
giving the indices of the integral being reduced,
and the integral appearing in its reduction.
The entry values can be fully numerical, if using rational
reconstruction methods to obtain the dependence on the
dimensional regulator $\eps$, or a series of coefficients
up to the required order in it.  The starting values
are likewise given by a sparse vector.  One then
carries out the reduction by multiplying a sequence
of sparse matrices.  With maximum ISP exponents
$(m_1,m_2)$, one expects to store $\Ord(m_1 m_2)$ 
sparse matrices (or equivalently, hash-table entries
for the recursive approach).  

While the optimal choice
of lowering operators is not yet clear, we can estimate
an upper bound on the number of operations required
for the reduction of top-level double-box integrals
by using the wall-hugging operator described in section \ref{DBBulkSection}.
Taking $m_1\sim m_2 \sim m$,
and ignoring constant factors, we see that $\Ord(m)$ 
applications of the wall-hugging operator would be 
required to get to a configuration with
$\powpropn_2 = 0$, at which point another $\Ord(m)$ 
applications of a boundary operator would be required to 
get to master integrals.  Each sparse matrix involved
may have up to $\Ord(m)$ terms, so that overall we will need 
$\Ord(m^2)$ operations for the reduction of the top-level
integrals.  For comparison, a naive Laporta approach
will start with $\Ord(m^2)$ different integrals, requiring 
Gaussian elimination on an $\Ord(m^2)\times\Ord(m^2)$ matrix.  
Without any tricks, this would require
$\Ord(m^6)$ operations.

\section{Conclusions}
\label{Conclusions}

In this article, we have shown how to construct a complete
set of lowering operators for Feynman integrals.  Each
application of such an operator reduces an integral
with arbitrary indices to simpler top-level and
daughter integrals.  A successive application of lowering
operators will reduce all top-level integrals to 
master integrals. In general these operators could also
be applied to integrals with non-integer indices. Similar operators can of course be
constructed for daughter integrals.  We studied two
examples, the planar double box in Sect.~\ref{DoubleBoxSection},
and the planar pentabox in Sect.~\ref{PentaboxSection}.
The required lowering operators for the
double box are given in 
eqs.~(\ref{BulkLoweringOperatorForm},%
\ref{RefinedBulkLoweringOperatorCoefficients},%
\ref{BoundaryLoweringOperatorForm},%
\ref{BoundaryLoweringOperatorCoefficients})
for undotted integrals,
and in the auxiliary
file \textsf{db\nobreakdash-operators.m\/} as well alongside
lowering operators for dotted integrals.  

Our construction generally yields operators with free
parameters.  These parameters can be chosen to enforce
special properties of the lowering operators. We have given an example of such a refinement in \eqn{BulkWallLoweringOperatorCoefficients}. We leave
an investigation of the optimal way to choose these parameters,
along with the optimal way to build lowering operators for
daughter integrals, to future investigation.

\begin{acknowledgements}
DAK thanks Giacomo Brunello, Fernando Febres Cordero, Bo Feng,
and Yang Zhang for helpful discussions and comments.
This work was supported by the European Research
Council (ERC) under the European Union’s research and
innovation program grant agreements ERC--AdG--885414 (`Ampl2Einstein').
\end{acknowledgements}

\appendix

\bibliographystyle{apsrev4-2mod}
\bibliography{main.bib}
        
\end{document}